\documentclass{cernyrep}
\usepackage{graphicx}
\usepackage{subfig}
\usepackage{multirow}
\usepackage{array}
\usepackage{amssymb}
\usepackage[T1]{fontenc}
\usepackage[bookmarks, colorlinks=true, linktoc=page, pdftex, linkcolor=black, citecolor=black, urlcolor=blue]{hyperref}
\usepackage{siunitx}
\usepackage{csquotes}
\usepackage{xargs}                      

\usepackage{tikz}
\usepackage{tikz-3dplot}
\usetikzlibrary{shapes,arrows}
\usetikzlibrary{positioning}
\tikzstyle{block} = [rectangle, draw, fill=blue!20, 
    text width=5em, text centered, rounded corners, minimum height=4em]
\tikzstyle{line} = [draw, -latex']

\begin{document}
\title{Particle Motion in Hamiltonian Formalism}
\author{Yannis Papaphilippou
                 }
\institute{CERN, Geneva, Switzerland}

\begin{abstract}
The goal of this contribution is to introduce the Hamiltonian formalism 
of theoretical mechanics for analysing motion in generic 
linear and non-linear dynamical systems, including particle accelerators. 
This framework allows the derivation and integration of equations of motion, 
in order to describe the particle trajectory evolution with respect to time. 
First, basic concepts are re-visited for describing particle propagation 
through the resolution of differential equations, applied
to linear and non-linear motion. These equations of motion can be obtained
by the Lagrangian of the system, which is the natural step 
leading to Hamiltonian Formalism and its properties. The accelerator ring Hamiltonian is derived, starting with the relativistic Hamiltonian of particles 
in the influence of E/M fields and a series of canonical 
(or symplectic) transformations and approximations. Thereby, introductory concepts 
of beam dynamics such as invariants and transport matrices are revisited and extended 
towards generic concepts such as action-angle variables and symplectic maps.
To this end, the ground is prepared for the advanced methods and tools used for studying 
non-linear motion in particle accelerators.
\end{abstract}


\maketitle

\section{Introduction}

The objective of these lecture notes is to provide an overview of methods for deriving and solving ("integrating") equations of motion, in order to describe the evolution, i.e. dependence with “time” of a system (“particle”). The formalism of theoretical (classical) mechanics will be introduced for analysing motion 
in general linear or non-linear dynamical systems, including particle accelerators. This will be the basis for connecting this formalism with concepts already explored in introductory level beam dynamics courses, 
such as  matrix solutions for analysing transverse motion, synchrotron motion, invariants, etc. Thereby, the ground is prepared for the advanced approaches for studying non-linear particle motion in accelerators.
Most of the material of this course is based on textbooks of classical mechanics~\cite{LandLif,Gold} and electromagnetism~\cite{Jackson}, books on non-linear dynamics and chaos with excellent introductory chapters~\cite{Tabor} but also recent books using as basis the Hamiltonian formalism to approach a wide range of aspects of beam dynamics in particle accelerators~\cite{Wolski, StuPen}.

At first, a reminder to the methods for resolving equations of motion is presented using as examples the simple harmonic oscillator and the pendulum. Then, the Lagrangian and Hamiltonian functions are introduced together with concepts of canonical variables and transformations, intimately connected to the preservation of phase-space volumes. At this point, the Hamiltonian for an accelerator ring can be obtained, applying a series of canonical transformations and approximations. Assuming linear fields,  the Hill's equations of betatron motion are re-derived and action-angle variables are introduced. Finally, symplectic maps as generalisation of linear matrices, are introduced.

\section{Equations of motion - the harmonic oscillator}
\label{sec:harm}
Equations of motion are necessary for describing the evolution
of a system of "particles" over time and determine their dynamics. In the simple
case of {\it 1 degree of freedom} (i.e. the position of the particle 
for any time $t$ is described by a unique variable $u$, like in planar motion), 
the differential equation describing the motion under a force $F$ due to a potential $V$ of
a particle with constant mass $m$ is given by Newton’s law:
\begin{equation}
m \ddot{u}  = m \frac{d^2u(t)}{dt^2} = F(u, \dot{u}, t) = - \frac{\partial V(u,\dot{u},t)}{\partial u} \;\;,
\label{eq:force}
\end{equation}
where the `` $\dot{}$ '' symbol represents the derivative over time, with the force $F$
and the potential $V$ being generally dependent on position $u$, velocity $\dot{u}$ and time $t$. Resolving or "integrating" this differential equation is essential for describing the motion of particles from
 initial conditions $u_0 = u(0)$, velocity $\dot{u_0}=\dot{u}(0)$ to an arbitrary phase space state 
 $(u(t),\dot{u}(t))$, at time $t$ and understanding the evolution of the system. 
 It can be easily shown by Taylor expanding any potential depending only in positions $V(u)$ 
 around an equilibrium point $u_0$, i.e. with vanishing first derivative $\displaystyle \frac{dV(u)}{du}\big{|}_{u=u_0}=0$, but finite second derivative $\displaystyle \frac{d^2V(u)}{du^2}\big{|}_{u=u_0}=k$, that the Newton's equation 
 can be approximated to the one of a simple harmonic oscillator with a linear restoring force, described by the differential equation:
\begin{equation}
\ddot{u} + \omega_0^2 u(t) = 0
\label{eq:harmosc}
\end{equation}
with the frequency $\displaystyle \omega_0= \sqrt{\frac{k}{m}}$, depending on the constant $k$ (e.g. the spring constant)
and the mass $m$. 

\begin{figure}[ht]
\begin{center}
 {\includegraphics[width=6cm]{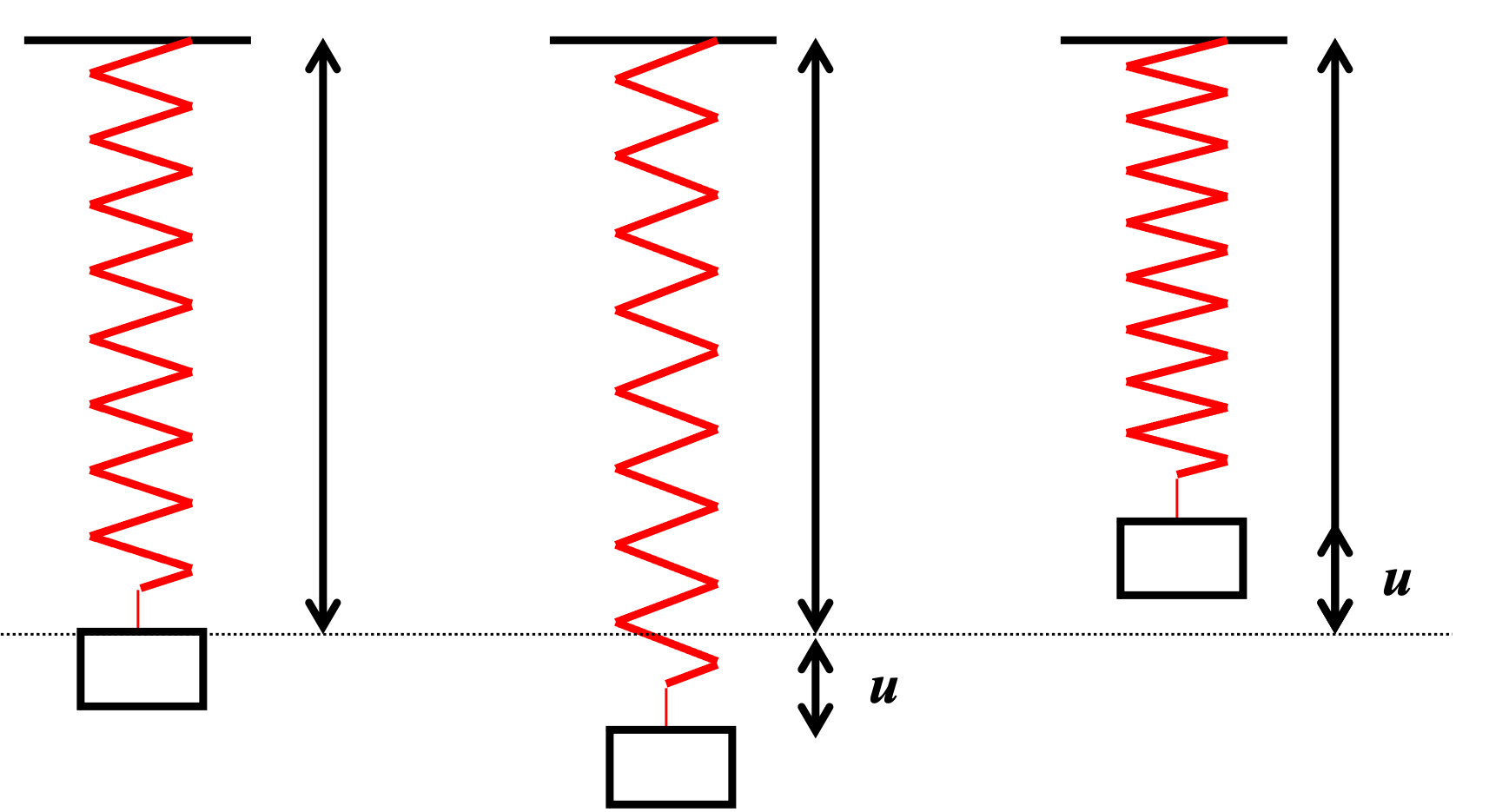}};
\caption{Schematic plot of a simple harmonic oscillator, of a spring with constant $k$ and mass $m$.}
\label{fig:harmosc}
\end{center}
\end{figure}

The solution is obtained by the {\it ansatz} substitution\footnote{More generically the solution can be considered to be a power series. The obtained recurrence relationship
of the series coefficients will end up to exactly the same result.}  $u=e^{\lambda t}$ and finding the solution of the characteristic polynomial  $\lambda^2+\omega_0^2 = 0$, i.e. $\lambda = \pm i\omega_0$ which yields the general solution 
 \begin{equation}
 u(t) = c e^{i\omega_0 t} + c^* e^{-i\omega_0 t}= C_1 \cos(\omega_0 t)+ C_2 \sin(\omega_0 t) = A \sin(\omega_0 t + \phi_0)\;\;,
 \label{harmsolu}
 \end{equation}
 with the ``velocity" 
\begin{equation}
  \frac{du(t)}{dt} = - C_1\omega_0 \sin(\omega_0 t)+ C_2\omega_0 \cos(\omega_0 t) = A\omega_0 \cos(\omega_0 t + \phi_0)\;\;.
  \label{harmsoldu}
\end{equation}
Note that a negative sign in the differential equation~\eqref{eq:harmosc} provides a solution described by hyperbolic sine/cosine functions. Note also that for no restoring force, for which $\omega_0 =0$, the motion is unbounded.

\subsection{Matrix solution}

The amplitude and phase depend on the initial conditions
\begin{equation}
u(0) = u_0 = C_1\;\;, \;\frac{du(0)}{dt} = \dot{u}_0 = C_2 \omega_0 \;\;, \; A=\frac{\left({\dot{u}_0}^2+\omega_0^2 u_0^2\right)^{1/2}}{\omega_0}\;\;, \; \tan(\phi_0) = \frac{\dot{u}_0}{\omega_0 u_0}
\end{equation}
The solutions can be re-written thus as
\begin{equation}
 \begin{aligned}
u(t) &= u_0 \cos(\omega_0 t)+ \frac{\dot{u}_0}{\omega_0}\sin(\omega_0 t) \;\;,\\
\dot{u}(t)  &= - u_0\omega_0 \sin(\omega_0 t)+ \dot{u}_0 \cos(\omega_0 t) \;\;.
\label{eq:solut}
\end{aligned}
\end{equation}
or in matrix form
\begin{equation}
\begin{pmatrix}
u(t) \\
\dot{u}(t) 
\end{pmatrix} = 
\begin{pmatrix}
\cos(\omega_0 t) & \frac{1}{\omega_0} \sin(\omega_0 t)\\
-\omega_0\sin(\omega_0 t) & \cos(\omega_0 t)\\
\end{pmatrix}
\begin{pmatrix}
u_0 \\
\dot{u}_0 
\end{pmatrix} \;\;.
\end{equation}
By replacing $\omega_0 \rightarrow \sqrt{k_0}$,	$t \rightarrow s$ and the dots with primes for the
``time" derivatives, this becomes the solution of a quadrupole~\cite{TraDyn}. 	
The general transfer matrix from position $s_0$ to $s$ is written as
\begin{equation}
\begin{pmatrix}
u \\
u' 
\end{pmatrix}_s = {\cal M}(s|s_0)
\begin{pmatrix}
u \\
u' 
\end{pmatrix}_{s_0} = 	
\begin{pmatrix}
C(s|s_0) & S(s|s_0) \\
C'(s|s_0) & S'(s|s_0)\\
\end{pmatrix}
\begin{pmatrix}
u \\
u' 
\end{pmatrix}_{s_0} \;\;.
\end{equation}
Note that the determinant of the matrix is $\det({\cal M}(s|s_0)) = C(s|s_0) S'(s|s_0) - S(s|s_0) C'(s|s_0)  = 1$, which is always true for conservative systems (``energy" is constant).
The general solution from $s_0$ to $s_n$ can be built by a series of matrix multiplications
\begin{equation}
{\cal M}(s_n|s_0) = {\cal M}(s_n|s_{n-1}) {\cal M}(s_{n-1}|s_{n-2}) \;\dots\; {\cal M}(s_2|s_1) {\cal M}(s_1|s_0)\;\;,
\end{equation}
as was already shown in transverse linear beam dynamics~\cite{TraDyn}.

\subsection{Integral of motion and integration by quadrature}
Another way to solve 2nd order differential equations is be splitting them 
to a pair of coupled first order equations. The equation for the harmonic 
oscillator then becomes
\begin{eqnarray}
\frac{du (t)}{dt} & = &p_u(t) \\
\frac{dp_u(t)}{dt}& = & -\omega_0^2 u(t) \;\;,
\end{eqnarray}
with $p_u$ the generalised momentum. By dividing the two sides of the equations, they can be combined to provide
\begin{equation}
\frac{d p_u}{dt} p_u +\omega_0^2 u \frac{du}{dt} = \frac{1}{2}\frac{d}{dt}\left (  p_u^2 +\omega_0^2 u^2 \right)  = 0 \;\;.
\end{equation}
By integrating, we get 
\begin{equation}
\frac{1}{2} \left(  p_u^2 +\omega_0^2 u^2 \right)  = I_1 \;\;,
\label{harmosc}
\end{equation}
 where $I_1$ is a constant or  {\it integral of motion} identified as the mechanical energy of the system. This equation describes in general an ellipse in phase space 
$(u,p_u)$. 	
Solving the previous equation for $\displaystyle p_u= \frac{du (t)}{dt}$, the system can be reduced to a first order equation 
\begin{equation}
\frac{du}{dt}  = \sqrt{2 I_1 - \omega_0^2 u^2 }\;\;.
\end{equation}
The last equation can be solved as an explicit integral or “quadrature”
\begin{equation}
\int dt = \int \frac{du}{\sqrt{2 I_1 - \omega_0^2 u^2 }}\;\;,
\end{equation}
yielding $t + I_2 =  \frac{1}{\omega_0} \arcsin\left( \frac{u\omega_0}{\sqrt{2I_1}}\right)$ or the well-known solution 
\begin{equation}
u(t) = \frac{\sqrt{2I_1}}{\omega_0}\sin(\omega_0 t +\omega_0 I_2)\;\;.
\end{equation}
Note that although the previous route may seem complicated, it becomes more natural when non-linear terms appear, where an ansatz of the type 	$u(t) = e^{\lambda t}$ is not applicable. The ability to integrate a differential equation is not just a nice mathematical feature, but deeply characterises the dynamical behaviour of the system described by the equation.

The period of the harmonic oscillator is calculated through the previous integral after integration between two extrema (when the velocity	$\displaystyle \frac{du}{dt}  = \sqrt{2 I_1 - \omega_0^2 u^2 }$ vanishes), i.e. $\displaystyle u_\text{ext}=\pm\frac{\sqrt{2I_1}}{\omega_0}
$:
\begin{equation}
T = 2\int_{-\frac{\sqrt{2 I_1}}{\omega_0}}^{\frac{\sqrt{2 I_1}}{\omega_0}} \frac{du}{\sqrt{2 I_1 - \omega_0^2 u^2 }} = \frac{2\pi}{\omega_0}\;\;.
\end{equation}
The period (or the frequency) of linear systems is independent of the integral of motion (energy). Note that this is not true for non-linear systems, e.g. for an oscillator with a non-linear restoring force, $$\frac{d^2u(t)}{dt^2} + k\; u(t)^3 = 0
\;\;.$$ In that case, the integral of motion is $I_1 = \frac{1}{2}p_u^2 +\frac{1}{4}k\;u^4 $ and the integration yields:
\begin{equation}
T = 2 \int_{-(4I_1/k)^{1/4}}^{(4I_1/k)^{1/4}} \frac{du}{\sqrt{2I_1-\frac{1}{2} k\; u^4}} = \sqrt{\frac{1}{2\pi }}\Gamma^2(\frac{1}{4})  \left({I_1\; k}\right)^{-1/4}
\;\;.
\end{equation}
This means that the period (frequency) depends on the integral of motion (energy), i.e. the maximum “amplitude” of the particle.

\section{A non-linear oscillator example: the pendulum}
\begin{figure}[ht]
\begin{center}
 {\includegraphics[width=4cm]{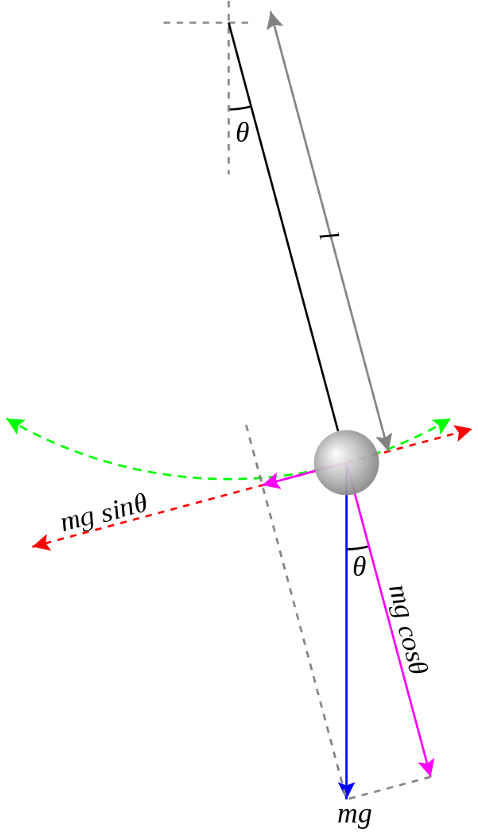}}
\caption{Schematic view of a pendulum.}
\label{fig:pend}
\end{center}
\end{figure}

An important non-linear equation which can be integrated is the one of the pendulum (see Fig.~\ref{fig:pend}), for a string of length $L$ and gravitational constant $g$
\begin{equation}
\frac{d^2\theta}{dt^2} + \frac{g}{L} \sin\theta = 0
\;\;.
\end{equation}
For small displacements, it reduces to a harmonic oscillator with frequency
$\displaystyle \omega_0 = \sqrt{ \frac{g}{L}}$.
By appropriate substitutions, this becomes the equation of synchrotron motion~\cite{LonDyn}.
The integral of motion (scaled energy) is 
\begin{equation}
\frac{1}{2}\left(\frac{d\theta}{dt}\right)^2 - \frac{g}{L} \cos\theta = I_1=E'
\end{equation}
and the quadrature is written as 
\begin{equation}
t = \int \frac{d\theta}{\sqrt{2(I_1+\frac{g}{L} \cos\theta)}}\;\;,
\end{equation}
assuming that for $t=0$,\;$\theta_0=\theta(0)=0$.
Using the substitutions	$ \cos\theta = 1-2 k^2\sin^2\phi$, with	$k = \sqrt{1/2(1+I_1 L/g)}$,  the integral is written as 
\begin{equation}
 t = \sqrt{\frac{L}{g}}\int_0^\theta \frac{d\theta}{\sqrt{1-k^2\sin^2\theta}} 
\end{equation}
and can be solved using the Jacobi elliptic sine function $\text{sn}$:
\begin{equation}
\theta(t) = 2\arcsin\left[k\; \text{sn}\left(t\sqrt{\frac{g}{L}},k\right)\right]\;\;.
\end{equation}

\begin{figure}[ht]
\begin{center}
 {\includegraphics[width=8cm]{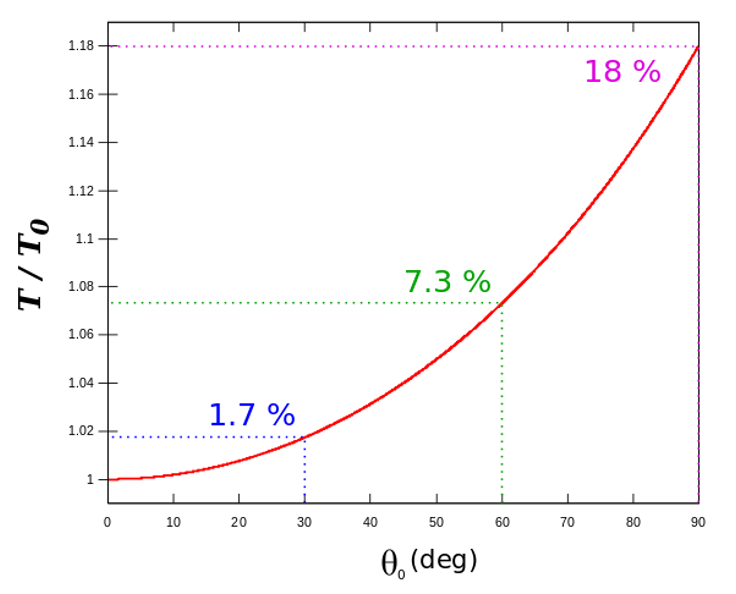}}
\caption{Ratio of the period of the pendulum with respect to the one of the harmonic oscillator, as a function of the pendulum angle.}
\label{fig:perpend}
\end{center}
\end{figure}

 For recovering the period, the integration is performed between the two extrema, i.e.               and $\theta=0$ and $\theta=\arccos(-I_1 L/g)$, corresponding to	$\phi = 0$ and $\phi=\pi/2$. The period is obtained by solving the integral
\begin{equation}
 T = 4\sqrt{\frac{L}{g}}\int_0^{\pi/2} \frac{d\phi}{\sqrt{1-k^2\sin^2\phi}} = 4\sqrt{\frac{L}{g}} {\cal K}(k) \;\;,
\end{equation}
i.e. it is given by the complete elliptic integral multiplied by four times the period of the harmonic oscillator. By expanding the complete elliptic integral $$
{\displaystyle {\cal K}(k)={\frac {\pi }{2}}\sum _{n=0}^{\infty }\left({\frac {(2n)!}{2^{2n}(n!)^{2}}}\right)^{2}k^{2n} = {\frac {\pi }{2}}\left(1+{\frac {1}{4}}k^{2}+\cdots \right)}
$$
with $k = \sqrt{1/2(1+I_1 L/g)}$, the “amplitude” dependence of the frequency becomes apparent.   
The deviation from the linear approximation becomes important at large amplitudes, as can be observed in Fig.~\ref{fig:perpend}. This dependence or spread of frequencies for beam particles with different amplitude is useful for damping coherent instabilities.

\section{Lagrangian and Hamiltonian functions}

\subsection{Lagrangian function}
 One of the main objectives of theoretical mechanics is to obtain equations for describing the motion of particles in $q_n$ coordinates for a system with $n$ degrees of freedom, from time $t_1$ to time $t_2$. This can be achieved by the {\it Lagrangian function} $L(q_1,\dots,q_n,\dot{q_1},\dots,\dot{q_n},t)$, with	$(q_1,\dots,q_n)$, the generalised coordinates and $(\dot{q_1},\dots,\dot{q_n})$ the generalised velocities. The Lagrangian is defined as $L=T-V$, i.e. the difference between kinetic and potential energy. 
The integral $$
{\cal S} = \int L(q_i,\dot{q_i},t)dt \;\;
$$	
defines the action. The principle of ``stationary action" or {\it Hamilton’s principle} states that any system evolves so as the action is stationary (its variation for small perturbation of the positions is zero), i.e. the action functional becomes extremum.
 By using Hamilton’s principle over some time interval between $t_1$ and $t_2$ for two stationary points $  \delta q(t_1) =\delta q(t_2) =0 $, the following differential equations, called the {\it  Euler-Lagrange equations}, for each degree of freedom are obtained (see appendix):
\begin{equation}
\frac{d}{dt}\frac{\partial L}{\partial \dot{q_i}} - \frac{\partial L}{\partial {q_i}} =0\;\;.
\end{equation}
In other words, by knowing the form of the Lagrangian, the equations of motion can be derived.

For a simple force law contained in a potential function governing the motion among interacting particles, the (classical) Lagrangian is (or as Landau-Lifshitz put it “experience has shown that…”~\cite{LandLif}):
\begin{equation}
L=T-V = \sum^n_{i=1} \frac{1}{2} m_i \dot{q}_i^2 - V(q_1,\dots,q_n)\;\;.
\label{eq:lang}
\end{equation}
 For velocity independent potentials, the Lagrange equations become
\begin{equation}
m_i \ddot{q_i} = - \frac{\partial V}{\partial q_i}\;\;,
\end{equation}
i.e. the well-known Newton’s equations.

\subsection{Hamiltonian formalism}
Although through the Lagrangian function, the goal of deriving equations of motions is achieved, there are certain disadvantages of the underlying formalism: First, the Lagrangian is not unique, in the sense that different Lagrangians can lead to the same equations. Second, the physical significance of the Lagrangian function is not 
straightforward. Even its basic form is given more by “experience” and the fact that it actually works that way! On the other hand, the Lagrangian has also certain advantages, as its relativistic form is very useful in particle physics, exploiting the fact that it is invariant under Lorentz transformations.

We have observed that the Lagrangian function provides in general  $n$ second order differential equations in the coordinate (or configuration) space. We have already shown though that it is quite advantageous to move to a system of $2n$ first order differential equations describing motion in phase space, which are more straightforward to solve. They can be derived by the Hamiltonian of the system, which is defined as the {\it Legendre transformation} of the Lagrangian      				
\begin{equation}
H({\bf q},{\bf p},t) =  \sum_i \dot{q}_i p_i - L({\bf q},{\bf\dot{q}},t)\;\;,
\end{equation}
where the generalised momenta are $\displaystyle p_i = \frac{\partial L}{\partial \dot{q}_i}$.
The generalised velocities can be  expressed as a function of the generalised momenta if the previous equation is invertible, and thereby define the Hamiltonian of the system.
As an example, let us consider the Lagrangian of \eqref{eq:lang}. From there, the momentum can be determined as $\displaystyle p_i = \frac{\partial L}{\partial \dot{q}_i} = m \dot{q}_i $, which can be trivially inverted to provide the well-known classical Hamiltonian of motion  
\begin{equation}
H({\bf q},{\bf p }) = \sum_i \frac{p_i^2}{2 m_i} + V(q_1,\dots,q_n)\;\;,
\end{equation}
which is the sum of kinetic and potential energy. The equations of motion or {\it Hamilton's equations} can be derived from the Hamiltonian following the same variational principle as for the Lagrangian (“stationary” action) but also by simply taking the differential of the Hamiltonian (see appendix):
\begin{equation}
\dot{q}_i = \frac{\partial H}{\partial p_i}\;, \;\; \dot{p}_i = - \frac{\partial  H}{\partial q}\;, \; \; \frac{\partial L}{\partial t} = -\frac{\partial H}{\partial t} 
\;\;.
\end{equation}
These are indeed $2n+2$ equations describing the motion in the “extended” phase space $(q_1,\dots,q_n,p_1,\dots,p_n,t,-H)$, as the Hamiltonian (as the Lagrangian) can present generally an explicite time dependence. These variables are called {\it canonically conjugate} (or canonical) and define the evolution of the system in phase space.
 They have the special property that they preserve volume in phase space, i.e. satisfy the well-known Liouville’s theorem. The variables used in the Lagrangian do not necessarily have this property.

Hamilton’s equations can be written in vector form 	$${\bf\dot{z}} = {\bf S} \cdot {\bf \nabla} H({\bf z})\;\;,$$ with ${\bf z}=(q_1,\dots,q_n,p_1,\dots,p_n)$ and ${\bf \nabla}=(\partial q_1,\dots,\partial q_n,\partial p_1,\dots,\partial p_n)$. The $2n\times 2n$ matrix $${\bf S} = \begin{pmatrix}
\;\; \bf{0}  & {\bf I} \\   - {\bf I} & \bf{0} 
\end{pmatrix}$$ is called the {\it symplectic} matrix.

\subsection{Poisson brackets}

A crucial step in the study of Hamiltonian systems is the identification of integrals of motion. Consider  a time dependent function of phase space. Its time evolution is given by 
\begin{equation}
\frac{d}{dt} f({\bf p},{\bf q},t) = \sum_{i=1}^n \left( \frac{dq_i}{dt}\frac{\partial f}{\partial q_i} + \frac{dp_i}{dt}\frac{\partial f}{\partial p_i} \right) + \frac{\partial f}{\partial t}
\end{equation}
Using Hamilton's equations, the sum can be re-written as
\begin{equation}
\frac{d}{dt} f({\bf p},{\bf q},t)  = \sum_{i=1}^n \left( \frac{\partial H}{\partial p_i} \frac{\partial f}{\partial q_i}  - \frac{\partial  H}{\partial q_i} \frac{\partial f}{\partial p_i} \right) + \frac{\partial f}{\partial t} = [H,f] + \frac{\partial f}{\partial t}\;\;,
\label{eq:pois}
\end{equation}
 where $[H,f]$  is the Poisson bracket of  $f$   with $H$. Some interesting properties of the Poisson bracket operators are given in the appendix.
If a quantity is explicitly time-independent and its Poisson bracket with the Hamiltonian vanishes (i.e. {\it commutes} with $H$), it is a constant (or integral) of motion, as the {\it autonomous} Hamiltonian itself.

\section{Canonical transformations}

It is very useful for some problems to find a set of variables that simplify the equations of motion, exploiting for example certain phase space symmetries. In this respect, it would be necessary to find a function $F_i$ for transforming the Hamiltonian from the variables $({\bf q},{\bf p})$ to $({\bf Q},{\bf P})$, while preserving the form of Hamilton's equation. When the transformation preserves the Hamiltonian properties (phase-space volume), it is called {\it canonical} ot {\it symplectic}. By mixing the old and new variables, there are four possible {\it generating} functions, which can be derived by~\cite{Gold}: 
\begin{align}
F_1({\bf q},{\bf Q}) :\;  & p_i  =  \;\;\frac{\partial F_1}{\partial q_i}, \;\, P_i  = -  \frac{\partial F_1}{\partial Q_i}  \label{eq:gener1}\\
F_2({\bf q},{\bf P}) :\;  & p_i  =  \;\;\frac{\partial F_2}{\partial q_i},\;\, Q_i  =    \;\;\frac{\partial F_2}{\partial P_i}\\
F_3({\bf Q},{\bf p}) :\;  & q_i  =  -\frac{\partial F_3}{\partial p_i}, \;\, P_i = -   \frac{\partial F_3}{\partial Q_i}  \\
F_4({\bf p},{\bf P}) :\;  & q_i  =  -\frac{\partial F_4}{\partial p_i}, \;\, Q_i =    \;\;\frac{\partial F_4}{\partial P_i}\;\;.
\label{eq:gener}
\end{align}
A general non-autonomous Hamiltonian is transformed to 
$$H({\bf Q}, {\bf P},t) = H({\bf q}, {\bf p},t) + \frac{\partial F_j}{\partial t}\;,\;\; j=1, 2, 3, 4\;\;.$$
One generating function can be constructed by using the other through Legendre transformations, e.g. 		         
\begin{equation}
F_2({\bf q},{\bf P}) =  F_1({\bf q},{\bf Q}) - {\bf Q \cdot P}\;,\;\; F_3({\bf Q},{\bf p}) =  F_1({\bf q},{\bf Q}) - {\bf q \cdot p} \;,\;\;\dots
\end{equation}
with the inner product defined as              			
\begin{equation}
{\bf q \cdot p} = \sum_i q_ip_i\;\;.
\end{equation}

\subsection{Preservation of Phase Volume}
A fundamental property of canonical transformations is the preservation of phase space volume (see appendix for a demonstration). This volume preservation in phase space can be represented in the old and new variables as
\begin{equation}
\int \prod_{i=1}^n dp_i dq_i = \int \prod_{i=1}^n dP_i dQ_i \;\;.
\end{equation}
The volume elements in old and new variables are related through the Jacobian
\begin{equation}
\prod_{i=1}^n dp_i dq_i = \frac{\partial (P_1,\dots,P_n,Q_1,\dots,Q_n)}{\partial (p_1,\dots,p_n,q_1,\dots,q_n)} \prod_{i=1}^n dP_i dQ_i \;\;.
\end{equation}
These two relationships imply that the Jacobian of a canonical transformation should have determinant equal to 1:
\begin{equation}
\left|\frac{\partial (P_1,\dots,P_n,Q_1,\dots,Q_n)}{\partial (p_1,\dots,p_n,q_1,\dots,q_n)} \right | = \left| \frac{\partial (p_1,\dots,p_n,q_1,\dots,q_n)}{\partial (P_1,\dots,P_n,Q_1,\dots,Q_n)} \right | = 1\;\;.
\end{equation}

\subsection{Examples of transformations}
The transformation $Q=-p\;, \;\; P=q$, which interchanges conjugate variables is area preserving, as the Jacobian is  
\begin{equation}
\frac{\partial (P,Q)}{\partial (p,q)} = 
\left |
\begin{matrix} 
\frac{\partial P}{\partial p} & \frac{\partial Q}{\partial p} \\
\frac{\partial P}{\partial q} & \frac{\partial Q}{\partial q}
\end{matrix} 
\right |
= 
\left |
\begin{matrix} 
0 & -1 \\
1 & \;\; 0\end{matrix} 
\right | =1 \;\;.
\end{equation}
On the other hand, the transformation from Cartesian to polar coordinates			$q=P \cos Q\;,\;\; p=P\sin Q$ is not, since
\begin{equation}
\frac{\partial (q,p)}{\partial (Q,P)} = 
\left |
\begin{matrix} 
-P\sin Q & P \cos Q\\
\cos Q &  \sin Q
\end{matrix} 
\right |
= -P
\;\;.
\end{equation}
There are actually “polar” coordinates that are canonical, given by	$q=-\sqrt{2P} \cos Q\;,\;\; p=\sqrt{2P}\sin Q$ for which
\begin{equation}
\frac{\partial (q,p)}{\partial (Q,P)} = 
\left |
\begin{matrix} 
\sqrt{2P}\sin Q & \sqrt{2P} \cos Q\\
-\frac{\cos Q}{\sqrt{2P}}& \frac{\sin Q}{\sqrt{2P}}
\end{matrix} 
\right |
= 1
\;\;.
\end{equation}

\section{The relativistic Hamiltonian for electromagnetic fields}

Neglecting self-fields and radiation, motion can be described by a “single-particle” Hamiltonian~\cite{Jackson}
\begin{equation}
H({\bf x},{\bf p}, t) = c \sqrt{\left({\bf p}-\frac{e}{c} {\bf A}({\bf x},t)\right)^2 + m^2c^2} + e \Phi({\bf x},t)\;\;,
\end{equation}
with ${\bf x} = (x,y,z)$ the Cartesian positions, ${\bf p} = (p_x,p_y,p_z)$ 
the conjugate momenta, ${\bf A} = (A_x,A_y,A_z)$ the magnetic vector potential
and $\Phi$ the electric scalar potential. The ordinary kinetic momentum vector is written as
$${\bf P} = \gamma m {\bf v } = {\bf p}-\frac{e}{c} {\bf A}\;\;,$$
with  ${\bf v }$ the velocity vector and $\gamma$ the relativistic factor.
The single-particle relativistic Hamiltonian is generally a 3 degrees of freedom one plus time (i.e., 4 degrees of freedom). The Hamiltonian represents the total energy
$$
H \equiv  E = \gamma m c^2 + e \Phi\;\;.
$$
The total kinetic momentum is $$
P = \left( \frac{H^2}{c^2} - m^2c^2 \right)^{1/2}\;\;.
$$
Using Hamilton's equations
$$
({\bf\dot{x}},{\bf\dot{p}}) = [({\bf x},{\bf
p}),H]\;\;,
$$
it can be shown that motion is governed by the Lorentz force equations~\cite{Jackson}.

\subsection{Hamiltonian for an accelerator ring}

\begin{figure}[ht]
\begin{center}
 {\includegraphics[width=10cm]{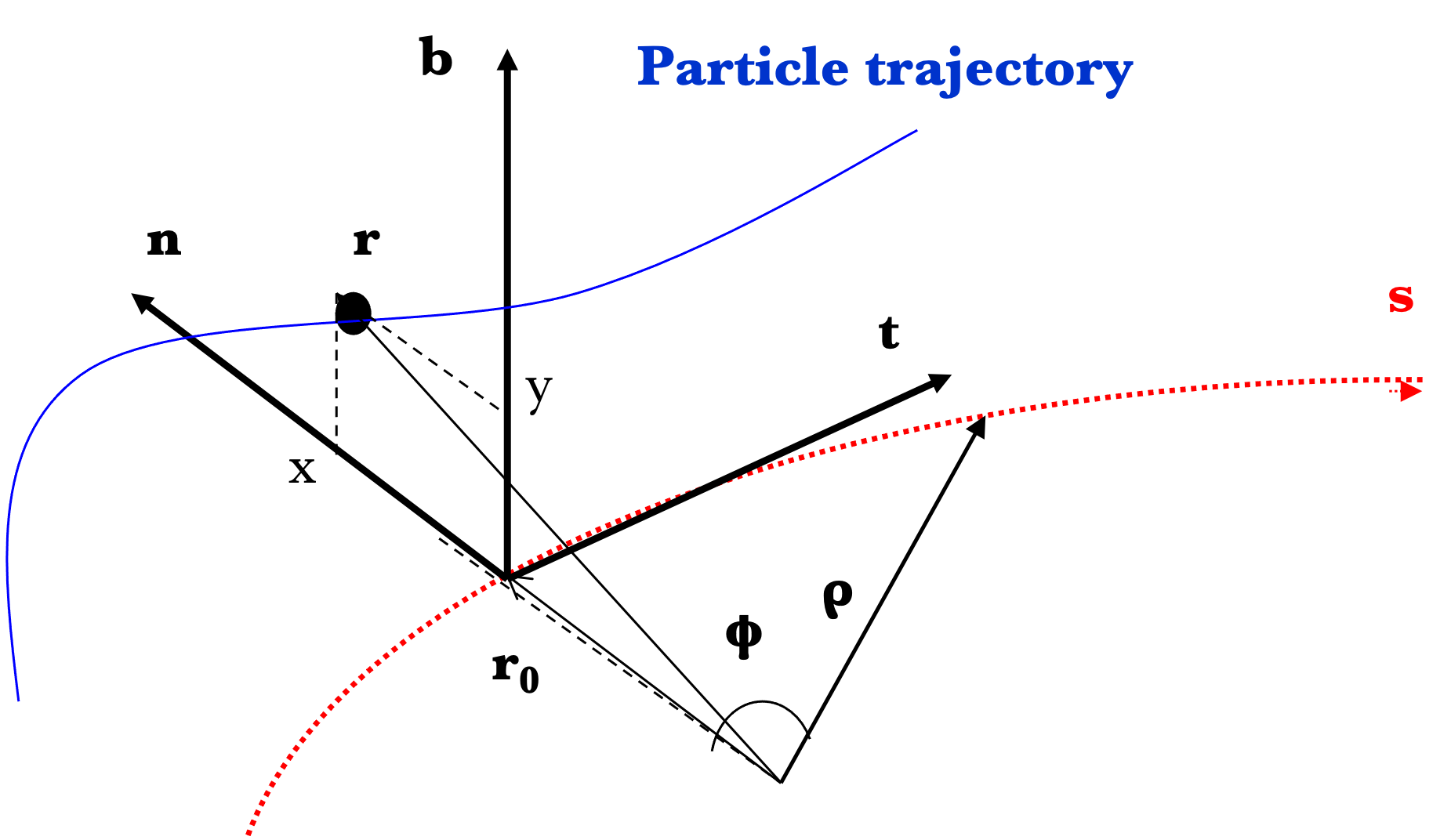}}
\caption{Frenet-Serret coordinate system.}
\label{fig:frenetserret}
\end{center}
\end{figure}

A series of canonical transformations and approximations are applied for simplifying the Hamiltonian for an accelerator ring. All derivations can be found in the appendix. Below we simply list them:
\begin{itemize} 
\item Move from Cartesian to Frenet-Serret rotating coordinate system $(x,y,z,p_x,p_y,p_z)  \mapsto  (X,Y,s,P_x,P_y,P_s)$, see Fig.~\ref{fig:frenetserret}, with bending in the horizontal plane which is useful for rings.
\item Change the independent variable from time $t$ to the path length $s$. The Hamiltonian can be considered as having 4 degrees of freedom, where the 4th “position” is time with conjugate momentum $P_t = -{\cal H}$ or $P_s = -{\cal H}$.
\item The electric field can be set to zero, as the longitudinal (synchrotron) motion is usually much slower than the transverse (betatron) one. 
\item Consider static and transverse magnetic fields.
\item Rescale the momentum with the reference one and move the origin to the periodic orbit.
\item Make approximations for the ultra-relativistic limit $\beta_0 \rightarrow 1\;, \;\; \frac{1}{\beta_0^2\gamma^2}\rightarrow 0$.
\end{itemize} 
The Hamiltonian then becomes 	
\begin{equation}
{\cal {H}} ({x}, {y}, l, {p}_x, {p}_y, \delta)=  (1+\delta)- e\hat{A}_s- \left(1+\frac{{x}}{\rho(l)}\right)
\sqrt{(1+\delta)^2 - {p}_x^2 - {p}_y^2}\;\;,
\label{eq:accHam}
\end{equation}								         
with the length $l = -ct+\frac{s-s_0}{\beta_0}$ and momentum deviation $ \frac{P_t-P_0}{P_0} \equiv \delta$. 

\subsection{High-energy and large ring approximation}
 
It is useful for study purposes, especially for finding an “integrable” version of the Hamiltonian, to make an extra approximation. We consider that the transverse momenta rescaled to the reference momentum are considered to be much smaller than 1, i.e. the square root of Eq.~\eqref{eq:accHam} can be expanded. Considering also the large machine approximation $x<<\rho$, where we drop cubic terms, the Hamiltonian is simplified to  
\begin{equation}
{\cal {H}}  = \frac{p_x^2+p_y^2}{2(1+\delta)} - \frac{x(1+\delta)}{\rho(s)} - e\hat{A}_s 
\label{eq:accHamhigh}
\end{equation}
This expansion may not be a good idea, especially for low energy, small size rings.

\subsection{General non-linear accelerator Hamiltonian}
Consider the general expression of the the longitudinal component of the vector potential is (see appendix)
\begin{itemize}
\item in curvilinear coordinates (curved elements) 
$$A_s = (1+\frac{x}{\rho(s)}) B_0 {\frak Re} \sum\limits_{n=0}^\infty
\frac{b_n + i a_n}{n+1} (x+iy)^{n+1} \;\;,
$$
\item In Cartesian coordinates 
$$A_s =  B_0 {\frak Re} \sum\limits_{n=0}^\infty
\frac{b_n + i a_n}{n+1} (x+iy)^{n+1} \;\;,
$$
\end{itemize}
with the skew and normal multipole coefficients being written as
$$
a_n = \frac{1}{B_0 n!} \frac{\partial^n B_x}{\partial x^n} \bigg{|}_{x=y=0}\;\text{and}\;
b_n = \frac{1}{B_0 n!} \frac{\partial^n B_y}{\partial x^n} \bigg{|}_{x=y=0}\;\;.
$$
The general non-linear Hamiltonian can be then written as 
\begin{equation}
{\cal H}(x,y,p_x,p_y,s) = {\cal H}_0(x,y,p_x,p_y,s) + \sum_{k_x,k_y} h_{k_x,k_y}(s) x^{k_x}y^{k_y}\;\;,
\end{equation}						           
with the periodic functions $h_{k_x,k_y}(s) = h_{k_x,k_y}(s+C)$ are functions of the multi-pole coefficients.

Here are the Hamiltonians for some magnetic elements:
\begin{itemize}
\item Dipole: $\displaystyle{H = \frac{x\delta}{\rho} +\frac{x^2}{2\rho^2}+ \frac{p_x^2+p_y^2}{2(1+\delta)}\;\;,}$
\item Quadrupole: $\displaystyle{H = \frac{1}{2} k_1(x^2 - y^2) + \frac{p_x^2+p_y^2}{2(1+\delta)}\;\;,}$
\item Sextupole: $\displaystyle{H = \frac{1}{3} k_2(x^3 - 3xy^2) + \frac{p_x^2+p_y^2}{2(1+\delta)}\;\;,}$
\item Octupole: $\displaystyle{H = \frac{1}{4} k_3(x^4 - 6x^2y^2 +y^4) + \frac{p_x^2+p_y^2}{2(1+\delta)}}$\;\;,
\end{itemize}
where $\displaystyle k_n = b_n(s) \frac{e}{c P_0}$ are the normalised multipole coefficients normalised to the magnetic rigidity $\displaystyle B\rho=\frac{P_0 c}{e}$. 

\subsection{Linear magnetic fields}

Assume a simple case of linear transverse magnetic fields, 
\begin{equation}
\begin{split}
B_x =& \quad b_1(s) y  \\
B_y =&-b_0(s) +  b_1(s) x
\end{split}
\label{magfields}\;\;,
\end{equation}
with the main bending field $\displaystyle B_0\equiv b_0(s)=\frac{P_0 c}{e \rho(s)}$ measured in Tesla, the normalized quadrupole gradient $\displaystyle K(s) = b_1(s) \frac{e}{c P_0}
= \frac{b_1(s)}{B\rho}$ measured in $1/\text{m}^2$ and the magnetic rigidity 
 $\displaystyle B\rho=\frac{P_0 c}{e}$. The vector potential has only a longitudinal component,
 which in curvilinear coordinates is
\begin{equation}
B_x = -\frac{1}{1+\frac{x}{\rho(s)}}\frac{\partial A_s}{\partial y}\;, \;\;
B_y = \frac{1}{1+\frac{x}{\rho(s)}}\frac{\partial A_s}{\partial x}\;\;.
\end{equation}
The previous expressions can be integrated to give the vector potential
\begin{equation}
A_s(x,y,s) = \frac{P_0 c}{e} \left[-\frac{x}{\rho(s)} -\left
(\frac{1}{\rho(s)^2} + K(s) \right)\frac{x^2}{2} + K(s) \frac{y^2}{2} \right]  =  P_0 c \ \hat{A}_s (x,y,s)\;\;.
\end{equation}
The Hamiltonian for linear fields can be finally written as
\begin{equation}
{\cal {H}}  = \frac{p_x^2+p_y^2}{2(1+\delta)} - \frac{x\delta}{\rho(s)} + \frac{x^2}{2\rho(s)^2} + \frac{K(s)}{2} (x^2 - y^2)\;\;.
\label{linham}
\end{equation}
Using Hamilton’s equation we have
\begin{equation}
\begin{split}
\frac{dx}{ds}  = \frac{p_x}{1+\delta}\;, \;\; & \frac{dp_x}{ds}  = \frac{\delta}{\rho(s)} - \left(\frac{1}{\rho^2(s)} + K(s) \right ) x\;\;, \\
\frac{dy}{ds}  = \frac{p_y}{1+\delta}\;, \;\; & \frac{dp_y}{ds}  = K(s) y\;\;.
\end{split}
\end{equation}
They can be written as two second order uncoupled differential equations, 
i.e. Hill’s equations~\cite{TraDyn}
\begin{equation}
\begin{split}
x'' & +  \frac{1}{1+\delta}\left( \frac{1}{\rho(s)^2} +  
K(s) \right) x =   \frac{\delta}{\rho(s)} \\
y'' &-  \frac{1}{1+\delta} K(s) y =  0
\end{split}
\end{equation}
with the usual solution for $\delta = 0$ and $u=x,y$
\begin{equation}
\begin{split}
u(s) =& \sqrt{\epsilon_u \beta_u (s)} \cos\left( \psi_u(s) +\psi_{u0} \right)\;\;, \\
u'(s) =& \frac{du}{ds} =  - \sqrt{\frac{\epsilon_u}{\beta_u (s)}} \left( \sin\left( \psi_u(s) +\psi_{u0} \right) + \alpha_u \cos\left( \psi_u(s) +\psi_{u0} \right)\right)\;\;.
\end{split}
\label{eq:C-S}
\end{equation}

\section{Action-Angle variables}

There is a canonical transformation to some optimal set of variables which can simplify the phase-space motion, the {\it action-angle variables}.
The action vector ${\bf J} = (J_1,J_2,\dots J_n)$ for a $n$-degree of freedom system  is defined as the integral
$$
{\bf J} =\frac{1}{2\pi}\oint {\bf p} d{\bf q}\;\;,
\label{eq:act}
$$
over closed paths in phase space. It can be proven that an {\it integrable Hamiltonian}, i.e. a Hamiltonian which provides equations of motion that can be integrated, is written as a function of only the actions, i.e. $H_0 = H_0({\bf J})$. Then, Hamilton’s equations give
\begin{equation}
\begin{split}
\dot\phi_i &= \;\;\, \frac{\partial H_0({\bf J})}{\partial J_i} =\omega_i({\bf J}) \Rightarrow  \phi_i = \omega_i({\bf J}) t +\phi_{i0}  \\
\dot J_i   &= - \frac{\partial H_0({\bf J})}{\partial \phi_i} = 0 \Rightarrow  J_i = \text{const.}
\end{split}\;\;,
\end{equation}
with $i = 1,2,\dots n$.
This means that the actions are {\it integrals of motion} and the conjugate angles are evolving linearly with time, with constant frequencies which depend on the actions.
The actions define the surface of an {\it invariant torus}, topologically equivalent to the product of circles.

\subsection{Action-angle variables for the harmonic oscillator}

The Hamiltonian for the harmonic oscillator can be written as (see Eq.~\eqref{harmosc})
\begin{equation}
H (u,p_u) = \frac{1}{2} \left(  p_u^2 +\omega_0^2 u^2 \right) \;\;,
\label{hamharmosc}
\end{equation}
with the canonical position and momentum $(u,p_u)$. From the definition of the action~\eqref{eq:act}
\begin{equation}
J_u =  \frac{1}{2\pi}\oint p_u du = \frac{1}{2\pi} \oint \sqrt{2H-\omega_0^2 u^2} du = \frac{1}{\pi} \int_{-u_\text{ext}}^{u_\text{ext}} \sqrt{2H-\omega_0^2 u^2} du = \frac{H}{\omega_0}\;\;,
\end{equation}
with $\displaystyle u_\text{ext} = \frac{\sqrt{2H}}{\omega_0}$ the position extrema, obtained for $p_u=0$, as defined in Sec.~\ref{sec:harm}. Thus, the Hamiltonian
in the new variables is written as $$H(\phi_u,J_u) = \omega_0 J_u\;\;.$$ The phase is found by Hamilton's
equations as $$\dot{\phi_u} = \frac{\partial H(\phi_u,J_u)}{\partial J_u}=\omega_0$$ and hence 
$\displaystyle \phi_u = \omega_0 t + \phi_{u,0}$, whereas the action $$\dot{J_u} = - \frac{\partial H(\phi_u,J_u)}{\partial \phi_u} = 0 \;\;,$$ i.e. $\displaystyle J_u = \text{const.}$ an integral of motion.

Another way to calculate the action is through a canonical transformation using a generating function. 
First, we observe from the solution of the harmonic oscillator (see Eqs.~\eqref{harmsolu},\eqref{harmsoldu}) that $$p_u = -\omega_0 u \tan\left( \omega_0 t + \phi_{u,0}\right) = -\omega_0 u \tan\left(\phi_{u}\right)\;\;,$$
which is a relationship already connecting the phase $\phi_u$ with coordinates $(u,p_u)$.
From the first generating function $F_1 (u,\phi_u)$ of Eq.~\eqref{eq:gener}, we have that $$ p_0 =\frac{\partial F_1}{\partial u} = -\omega_0 u \tan\left(\phi_{u}\right)$$ and by integrating we obtain
the generating function 
$$F_1 = \int p dx = -\frac{\omega_0 u^2}{2} \tan(\phi_u)\;\;.$$ 
The new momentum $J_u$ conjugate to $\phi_u$ is obtained by 
$$J_u = - \frac{\partial F_1}{\partial \phi_u} = \frac{\omega_0 u^2}{2}(1+\tan^2({\phi_u})) = \frac{1}{2\omega_0}(\omega_0^2 u^2 + p^2) = \frac{H}{\omega_0}\;\;,$$
i.e. exactly the same relationship as with the aforementioned method.

\subsection{Accelerator Hamiltonian in action-angle variables}
Considering on-momentum motion, the Hamiltonian~\eqref{linham} can be written as
\begin{equation}
{\cal {H}}_0  = \frac{p_x^2+p_y^2}{2} + \frac{K_x(s) x^2- K_y(s) y^2}{2} \;\;,
\label{lindeltham}
\end{equation}
with $\displaystyle K_x (s) = \frac{1}{\rho(s)^2} + K(s)$ and $\displaystyle K_y(s) = K(s)$.
As for the harmonic oscillator, we can use the Courant-Snyder solutions~\eqref{eq:C-S} and build the generating function
$$
F_1(x,y,\phi_x,\phi_y;s) = -\frac{x^2}{2\beta_x(s)}\left[\tan(\phi_x(s)) +a_x(s) \right ] -\frac{y^2}{2\beta_y(s)}\left[\tan(\phi_y(s)) +a_y(s) \right ]\;\;.
$$
The old variables with respect to actions and angles are
$$u(s) =   \sqrt{2J_u\beta_u(s)}\cos(\phi_u(s))\;,\;\; p_u(s) =-\sqrt{\frac{2J_u}{\beta_u(s)}}
\left(\sin(\phi_u(s))
+\alpha_u(s)\cos(\phi_u(s))\right)\;\;,
$$
which is just the solution of the Hill's equations with the usual Courant-Snyder invariant being a simple function of the action $\epsilon_u = 2 J_u$ and $\phi_u (s) = \psi_u(s)+\psi_u(0)$. 
The Hamiltonian~\eqref{lindeltham} takes the form
\begin{equation}
{\cal {H}}_0 (J_x,J_y;s)  = \frac{J_x}{\beta_x(s)} + \frac{J_y}{\beta_y(s)}\;\;.
\label{Hamactan}
\end{equation}
A further transformation to normalised coordinates 
$$
\begin{pmatrix}
{\cal U}\\ 
{\cal P_U'}
\end{pmatrix}
= 
\begin{pmatrix}
\frac{1}{\sqrt{\beta_u}} & 0\\ 
\frac{\alpha_u}{\sqrt{\beta_u}} & \sqrt{\beta_u}
\end{pmatrix}
\begin{pmatrix}
u\\ 
p_u
\end{pmatrix}
\;\;
$$
provides the following
$$
\begin{pmatrix}
{\cal U}\\ 
{\cal P_U'}
\end{pmatrix}
= \sqrt{2J_u}
\begin{pmatrix}
\cos(\phi_u)\\ 
- \sin(\phi_u)
\end{pmatrix}\;\;,
$$
describing simple rotations in phase space. Still, the phase $\phi_u$ is not advancing linearly with ``time" (longitudinal position $s$), as in the case of proper action-angle variables, because the Hamiltonian is non-autonomous. Exploiting its periodicity, a 1-turn Hamiltonian can be obtained by integrating the Hamiltonian~\eqref{Hamactan} around one turn
\begin{equation}
\bar{\cal {H}}_0 (J_x,J_y)  = \omega_x J_x + \omega_y J_y \;\;.
\end{equation}
with $$\omega_u = 2\pi Q_u =\oint\frac{d u}{\beta_u(s)}\;\;. $$ The motion is the one of two linearly independent harmonic oscillators with frequencies the tunes $Q_u$.

\subsection{Linear normal forms}
The previous change of variables (action-angle and normalised) can be considered a transformation to linear {\it normal forms}. The idea is to find a coordinate transformation so that the motion becomes simpler. This transformation is meant to simplify the motion in phase space, e.g. in the case of betatron motion to transform ellipses to circles so that the motion is described by simple rotations through the matrix 
$$
{\cal R} = 
\begin{pmatrix}
\cos\phi & \sin\phi \\ 
-\sin\phi & \cos\phi 
\end{pmatrix}
\;\;.
$$ 
Considering the general matrix ${\cal M}(s)$ describing betatron motion
\begin{equation}
{\cal M}_s = \begin{pmatrix}
\sqrt{\frac{\beta(s)}{\beta_0}} \left( \cos\phi + \alpha_0\sin\phi \right) & \sqrt{{\beta(s)}{\beta_0}} \sin\phi  \\
\frac{\left(\alpha_0 - \alpha(s)\right) \cos\phi - \left(1+\alpha_0\alpha(s) \right) \sin\phi}{\sqrt{{\beta(s)}{\beta_0}}} & 
\sqrt{\frac{\beta_0}{\beta(s)}} \left( \cos\phi - \alpha_0\sin\phi \right)
\end{pmatrix}
\end{equation}
the transformation can be found through the matrix multiplication operation 
$$
{\cal M}(s) = {\cal T}(s)^{-1} \circ {\cal R} \circ {\cal T}(0) \Leftrightarrow   {\cal R} = {\cal T}(s) \circ {\cal M}_s \circ {\cal T}(0)^{-1}\;\;,
$$
which looks like a type of diagonalisation. After some algebra, we find that 
$$
{\cal T}(s) = 
\begin{pmatrix}
\frac{1}{\sqrt{\beta(s)}} & 0\\ 
\frac{\alpha(s)}{\sqrt{\beta(s)}} & \sqrt{\beta(s)}
\end{pmatrix}\;\;.
$$
This transformation can be extended to a non-linear system involving non-linear maps~\cite{non-dyn}.

\section{Symplectic maps}

\begin{figure}[ht]
\begin{center}
 {\includegraphics[width=8cm]{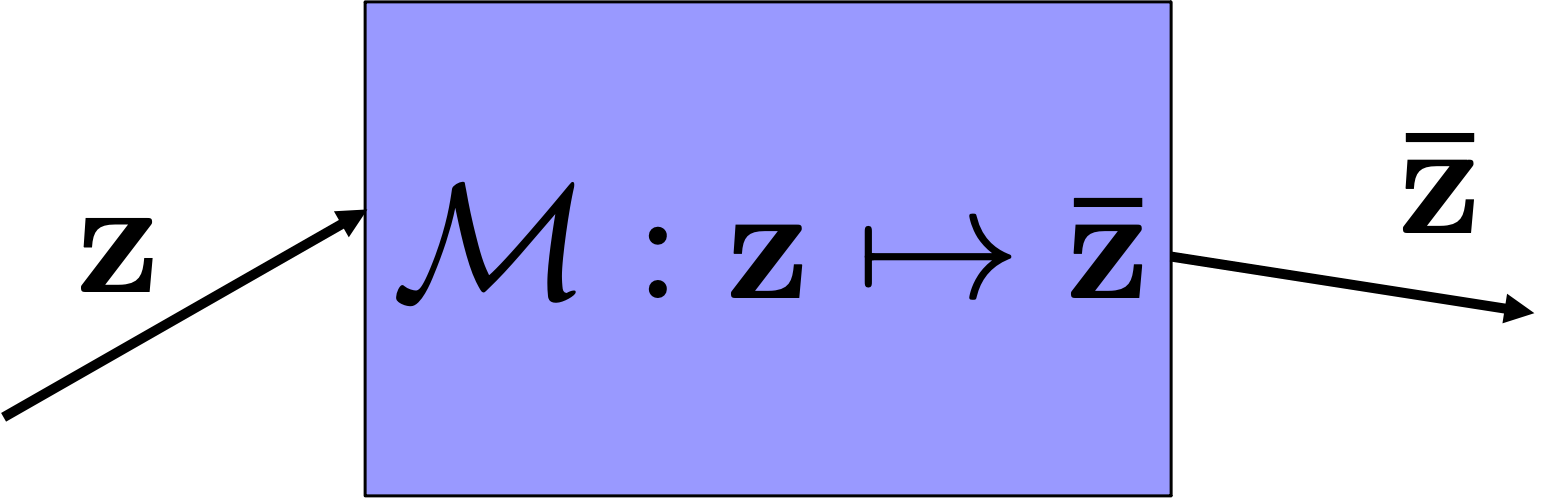}}
\caption{Schematic representation of a map.}
\label{fig:map}
\end{center}
\end{figure}

A generalization of the matrix (which can only describe linear systems) is a map ${\cal M}$ (see Fig.~\ref{fig:map}), which transforms a system from some initial to some final coordinates.
Analyzing the map will give useful information about the behaviour of the system.
There are different ways to build the map but always the preservation of symplecticity is important.

Consider two sets of canonical variables $\displaystyle {\bf z}$ and $\displaystyle {\bf\bar{z}}$, which may be even considered as the evolution of the system between two points in phase space.
A transformation from the one to the other set can be constructed through a map 
${\mathcal{M}}: {\bf z} \mapsto {\bf\bar{z}}$.
The Jacobian matrix of the map $M=M({\bf z},t)$ is composed by the elements $\displaystyle M_{ij} \equiv {\partial{\bar{z}_i}\over\partial{z_j} }
$
The map is symplectic if the Jacobian matrix satisfies the condition $\displaystyle M^T {\cal S} M = {\cal S}$ where $${\cal S} = 
\begin{pmatrix}
\;\;\;\bf{0} & \bf{I} \\
-\bf{I}& \bf{0}         
\end{pmatrix}\;\;.
$$
It can be shown that the determinant of the Jacobian matrix of symplectic map is $\det(M) = 1$.
It can be also shown that the Poisson brackets of the variables defined through a symplectic map $[{\bar{z}}_i,{\bar{z}}_j] =
[{z}_i,{z}_j] = {\cal I}_{ij}$,	which is a known relation satisfied by canonical variables.
In other words, symplectic maps {\it preserve} the Poisson brackets.
\begin{figure}[ht]
\begin{center}
 {\includegraphics[width=10cm, height=6cm]{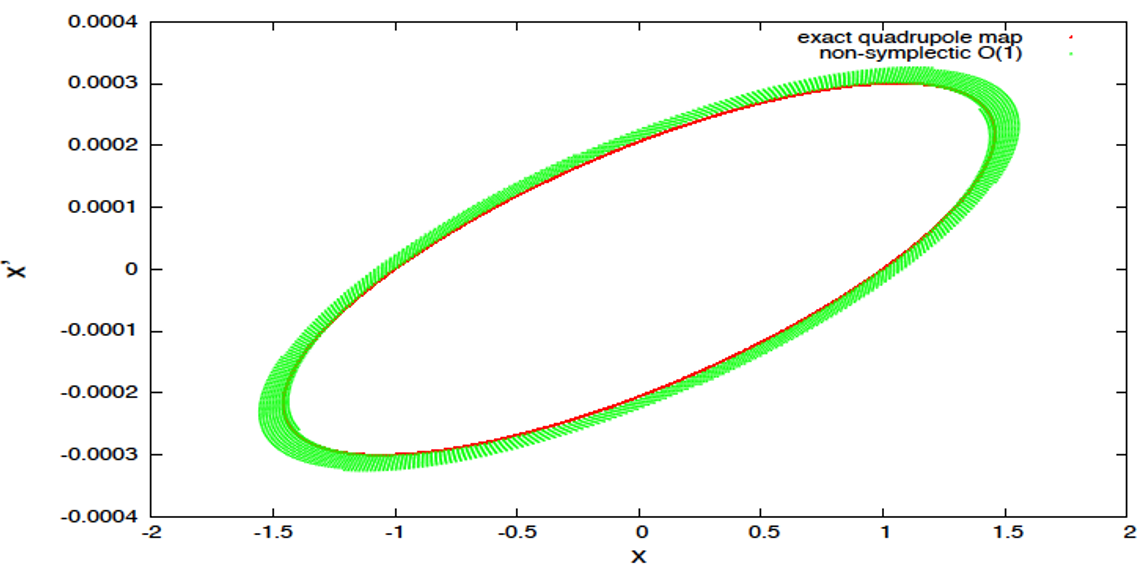}}
\caption{Phase space of the exact (red) and approximated (green) quadrupole matrix.}
\label{fig:phasp}
\end{center}
\end{figure}

Why symplecticity is so important? In fact, it guarantees that the transformations in phase space are area preserving. To understand what deviation from symplecticity produces consider the simple case of the quadrupole with the general matrix written as
$$
{\cal M}_\text{Q}  = 
\begin{pmatrix} \cos(\sqrt{k} L)  & \frac{1}{\sqrt{k}}\sin(\sqrt{k} L ) \\ -\sqrt{k} \sin(\sqrt{k} L )    & \cos(\sqrt{k} L) \end{pmatrix}\;\;.
$$
By taking the Taylor expansion for small lengths, up to first order, we obtain
$$
{\cal M}_\text{Q}  = 
\begin{pmatrix} 1 & L \\ -k L    & 1 \end{pmatrix} + O(L^2)\;\;.
$$
This matrix is indeed not symplectic as its determinant is equal to	$1+kL^2$, 
i.e. there is a deviation from symplecticity at 2nd order in the quadrupole length.
It can become though very close to symplectic if the length becomes infinitesimally small.

The iterated non-symplectic matrix in the phase space represented in Fig.~\ref{fig:phasp} does not provide the well-know elliptic trajectory (red elliptic curve versus green points). Although the trajectory is very close to the original one, it spirals outwards 
towards infinity, which obviously cannot be physically correct.

\section{Summary}
Starting from basic mathematical concepts, we have shown that second order differential equations of motion coming from from Newton’s law (configuration space) can be solved by transforming them to pairs of first order ones (in phase space). In that way, there is the natural appearance of invariant of motion  (“energy”). Extending this to non-linear oscillators,
we verified that their frequencies depend on the invariant (or “amplitude”). We thereby
connected the invariant of motion to the system’s Hamiltonian, derived through the Lagrangian.
Next we have shown that through the Hamiltonian, the equations of motions can be derived.
The Poisson bracket operators are helpful for discovering integrals of motion. At the same time,
canonical (or symplectic) transformations are necessary for preserving the phase-space volume. 
Using as starting point the relativistic Hamiltonian of particles in electromagnetic fields, and using 
a series of canonical transformations and approximations, the accelerator ring Hamiltonian was derived.
Imposing linear magnetic fields in the accelerator Hamiltonian, Hamilton’s equations provide the usual  Hill’s equation, thereby showing that the Hamiltonian formalism leads to the same physics.
The linear (uncoupled) magnetic field Hamiltonian can be simplified through transformation in action-angle variables, being only function of the actions. The motion in phase space can be transformed from ellipses to simple circles. Finally, symplectic maps are essential for preserving the correct physical time evolution of linear or non-linear systems.

\section{Appendix}
\subsection{Derivation of Lagrange equations}
The variation of the action can be written as
\begin{equation}
\delta W = \int^{t_2}_{t_1} \left( L(q+\delta q, \dot{q} + \delta\dot{q} , t) - L(q, \dot{q}  , t) \right) dt = \int^{t_2}_{t_1} \left( \frac{\partial L}{\partial q}\delta q + \frac{\partial L}{\partial \dot{q}} \delta \dot{q} \right )dt \;\;.
\end{equation}
Taking into account that $\displaystyle \delta \dot{q} = \frac{d(\delta q)}{dt}$, the 2nd part of the integral can be integrated by parts giving   
$$
\delta W = \left| \frac{\partial L}{\partial \dot{q}} \delta q \right|_{t_1}^{t_2 } + \int^{t_2}_{t_1}  \left( \frac{\partial L}{\partial q} - \frac{d}{dt} \left(\frac{\partial L}{\partial \dot{q}} \right) \right ) \delta q dt = 0 \;\;.
$$
The first term is zero because	$ \delta q(t_1) =\delta q(t_2) =0 $, so the second integrant should also vanish, providing the  following differential equations for each degree of freedom, the Lagrange equations
$$
\frac{d}{dt}\frac{\partial L}{\partial \dot{q_i}} - \frac{\partial L}{\partial q_i} = 0\;\;.
$$
\subsection{Derivation of Hamilton’s equations}
The equations of motion can be derived from the Hamiltonian following the same variational principle as for the Lagrangian (“least” action) but also by simply taking the differential of the Hamiltonian				d  $$
d H (q,p,t)  = \sum_i p_i  d\dot{q}_i + \dot{q}_i dp_i- \frac{\partial L}{\partial \dot{q}_i} d\dot{q}_i - \frac{\partial L}{\partial q_i} dq_i - \frac{\partial L}{\partial t}dt\;\;
$$
Using from Lagrange equations that $\displaystyle p_i= \frac{\partial L}{\partial \dot{q}_i}$ and $\displaystyle \dot{p_i} = \frac{\partial L}{\partial q_i}$, the previous equation is written
$$
d  H (q,p,t) = \sum_i \dot{q}_i dp_i- \dot{p_i} dq_i - \frac{\partial L}{\partial t}dt\;\;.
$$
But the differential of the Hamiltonian is also 
$$
d  H (q,p,t) = \sum_i \frac{\partial H}{\partial p_i} dp_i + \frac{\partial H}{\partial q_i} dq_i + \frac{\partial H}{\partial t}dt\;\;.
 $$
By equating terms multiplied by the same differentials, the Hamilton’s equations are derived
$$
\dot{q}_i = \frac{\partial H}{\partial p_i}\;, \;\; \dot{p}_i = - \frac{\partial  H}{\partial q}\;, \; \; \frac{\partial L}{\partial t} = -\frac{\partial H}{\partial t} \;\;.
$$
These are indeed $2n+2$	equations describing the motion in the “extended” phase space $(q_i,\dots,q_n,p_1,\dots,p_n,t,-H)$.

\subsection{Preservation of phase-space volume}
A fundamental property of Hamiltonian systems is the preservation of phase-space volume as they evolve.
Let’s have a system evolving from $(p_i q_i) \rightarrow (p'_i q'_i)$ after time $\delta t$. 
By Taylor-expanding and using Hamilton’s equations we have:
\begin{equation*}
\begin{split}
q'_i = q_i(t+\delta t) = q_i(t) + \frac{dq_i}{dt} \delta t + O(\delta t^2)  = q_i(t) - \frac{\partial H}{\partial p_i} \delta t + O(\delta t^2)\\
p'_i = p_i(t+\delta t) = p_i(t) + \frac{dp_i}{dt} \delta t + O(\delta t^2)  = p_i(t) + \frac{\partial H}{\partial q_i} \delta t + O(\delta t^2)
\end{split}\;\;.
\end{equation*}
By differentiating, we get
\begin{equation*}
\begin{split}
dq'_i = d q_i(t) - \frac{\partial}{\partial q_i} \left( \frac{\partial H}{\partial p_i}\right ) dq_i \delta t + O(\delta t^2)\\
dp'_i = d p_i(t) + \frac{\partial}{\partial p_i} \left( \frac{\partial H}{\partial q_i}\right ) dp_i \delta t + O(\delta t^2)
\end{split}\;\;.
\end{equation*}
Finally, by multiplying the two equations
$$
dq'_i dp'_i = d q_i d p_i \left [ 1 - \frac{\partial}{\partial q_i} \left( \frac{\partial H}{\partial p_i}\right ) +\frac{\partial}{\partial p_i} \left( \frac{\partial H}{\partial q_i}\right ) \right] \delta t + O(\delta t^2) \approx d q_i d p_i\;\;,
$$
which proves the preservation of the phase space volumes for canonical variables defining a Hamiltonian.
\subsection{Poisson bracket properties}

From the definition of Poisson brackets of Eq.~\eqref{eq:pois}, and for any three given functions, the following properties can be shown:
\begin{align*}
\displaystyle
&\text{bilinearity:}\quad [a f + b g,h] = a[f,h] + b[g,h]\;, a,b\in \Bbb{R} \\
&\text{anticommutativity:}\quad [f,g] = - [g,f] \\
&\text{Jacobi’s identity:}\quad [f,[g,h]] + [g,[h,f]] + [h,[f,g]] = 0  \\
&\text{Leibniz’s rule:}\quad [f,gh] = [f,g]h + g[f,h]\;\;. 
&\end{align*}
The first three of the previous properties prove that the Poisson brackets operation satisfies a {\it Lie algebra}.

\subsection{Magnetic multipole expansion}

\begin{figure}[ht]
\begin{center}
 {\includegraphics[width=6cm]{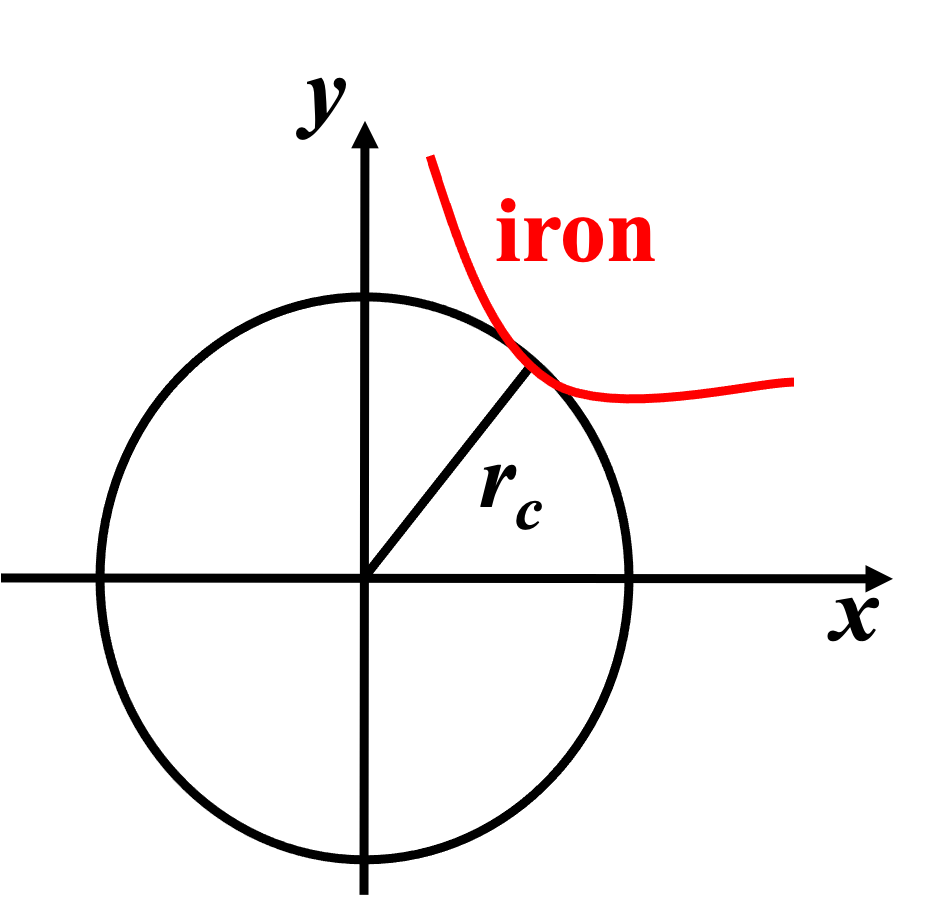}}
\caption{Schematic of the magnet coordinate system with the reference radius $r_c$ and the iron yoke.}
\label{fig:iron}
\end{center}
\end{figure}

From Gauss law of magnetostatics, $\displaystyle {\bf \nabla\cdot B = 0}$, a vector potential exists that 
${\bf B =  \nabla \times A }$. 
Assuming transverse 2-dimensional fields, the vector potential has only one longitudinal component $A_s$. 
The Ampere’s law in vacuum (inside the beam pipe) is ${\bf \nabla \times B = 0 }$, which means that the magnetic induction can be defined through a scalar potential ${\bf B} = - \nabla V$.
Using the previous equations, the relations between field components and potentials are
$$
B_x = - \frac{\partial V}{\partial x} = \frac{\partial A_s}{\partial y} \;\;\text{and}\;\;
B_y = - \frac{\partial V}{\partial y} = - \frac{\partial A_s}{\partial x} \;\;,
$$
i.e. Riemann conditions of an analytic function. This means that we can define a complex potential ${\cal A}(z)$ of a variable $z=x+iy$, the power series expansion of which is convergent in a circle with radius $r_c$(distance from iron yoke as shown in Fig.~\ref{fig:iron}). The expansion can be written
$$
{\cal A}(x+iy) = A_s(x,y)+iV(x,y) = \sum_{n=1}^{\infty} \kappa_n z^n = \sum_{n=1}^{\infty} (\lambda_n+i\mu_n) (x+iy)^n\;\;.
$$
From the complex potential, we can derive the magnetic fields
$$
B_y+i B_x = -\frac{\partial}{\partial x} \left( A_s(x,y)+iV(x,y) \right)  = - \sum_{n=1}^{\infty} n (\lambda_n+i\mu_n) (x+iy)^{n-1}\;\;.
$$
Setting	$b_n = -n \lambda_n$ and $a_n = n \mu_n$, we get the multi-pole expansion
$$
B_y+i B_x =  \sum_{n=1}^{\infty}  (b_n - i a_n) (x+iy)^{n-1}\;\;.
$$
We can thereby define normalised multi-pole coefficients
$\displaystyle b_n' = \frac{b_n}{10^{-4}B_0}r^{n-1}$ and $\displaystyle a_n' = \frac{a_n}{10^{-4}B_0}r^{n-1}$, on a reference radius $r_0$, $10^{-4}$ of the main field $B_0$ to get
$$
B_y+i B_x =  10^{-4} B_0 \sum_{n=1}^{\infty}  (b_n' - i a_n') \left(\frac{x+iy}{r_0}\right )^{n-1}\;\;.
$$
Note that the USA convention is obtained for $n' = n-1 $.

\subsection{Canonical transformations and approximations for the accelerator ring Hamiltonian}
\subsubsection{From Cartesian to “curved” coordinates}
 It is useful (especially for rings) to transform the Cartesian coordinate system to the                                       Frenet-Serret system moving to a closed curve, with path length $s$.
The position coordinates in the two systems are connected by
$\displaystyle {\bf r} = {\bf r_0}(s) + X {\bf n}(s) + Y {\bf b}(s) = x {\bf u_x} + y {\bf u_y} + z {\bf u_z}$.
The Frenet-Serret unit vectors are defined as
$
\displaystyle
({\bf t}, {\bf n}, {\bf b}) = (\frac{d}{ds}{\bf r_0}(s) , - \rho(s)\frac{d^2}{ds^2}{\bf r_0}(s), {\bf t} \times {\bf n} )
$
and their derivatives are given by
$$
\frac{d}{ds}
\begin{pmatrix}
{\bf t} \\ {\bf n} \\ {\bf b}
\end{pmatrix}
= 
\begin{pmatrix}
0 & - \frac{1}{\rho(s)} & 0 \\
\frac{1}{\rho(s)}  & 0 & - \tau(s) \\
0& 0 & \tau(s) 
\end{pmatrix}
\begin{pmatrix}
{\bf t} \\ {\bf n} \\ {\bf b}
\end{pmatrix}\;\;,
$$		  
with $\rho(s)$ the radius of curvature and $\tau(s)$ the torsion which vanishes in case of planar motion.
We are now seeking a canonical transformation between 
\begin{eqnarray*}({\bf q},{\bf p}) & \mapsto & ({\bf Q},{\bf P}) \;\; \text{or}\\
(x,y,z,p_x,p_y,p_z) & \mapsto & (X,Y,s,P_x,P_y,P_s)\;\,.\end{eqnarray*}
The generating function is 
$$
({\bf q},{\bf P}) = -(\frac{\partial F_3({\bf p},{\bf Q})}{\partial {\bf p}}, \frac{\partial F_3({\bf p},{\bf Q})}{\partial {\bf Q}} )\;\;.
$$
By using the relationship for the positions, 
the generating function is 					            							
$F_3({\bf p}, {\bf Q}) = -{\bf p \cdot  r} $.
For planar motion, the momenta are
$$
{\bf P} = (P_X,P_Y, P_s) = {\bf p\cdot} (\frac{\partial F_3}{\partial X}, \frac{\partial F_3}{\partial Y}, \frac{\partial F_3}{\partial s} ) = {\bf p\cdot} ({\bf  n}, {\bf  b}, (1+\frac{X}{\rho}) {\bf t} )\;\;.
$$
Taking into account that the vector potential is also transformed in the same way
$$
(A_X,A_Y,A_s) = {\bf A\cdot} ({\bf  n}, {\bf  b}, (1+\frac{X}{\rho}) {\bf t} )\;\;,
$$											         	         									         the new Hamiltonian is given by 
$$
{\cal H}({\bf Q},{\bf P},t) = c
\sqrt{(P_X-\frac{e}{c}A_X)^2 + (P_Y-\frac{e}{c}A_Y)^2+
\frac{(P_s-\frac{e}{c}A_s)^2}{(1+\frac{X}{\rho(s)})^2} + m^2c^2} + e
\Phi\;\;.
$$

\subsubsection{Changing of the independent variable}
 It is more convenient to use the path length  $s$, instead of the time as independent variable.
 The Hamiltonian can be considered as having 4 degrees of freedom, where the 4th “position” is time and its conjugate momentum is $P_t = -{\cal H}$.
In the same way, the new Hamiltonian with the path length as the independent variable is just
$P_s = -{\cal \tilde H}(X,Y,t,P_X,P_Y, P_t, s)$, with
$$
{\cal \tilde H} = - \frac{e}{c}A_s- \left(1+\frac{X}{\rho(s)}\right)
\sqrt{(\frac{P_t + e \Phi}{c})^2 - m^2 c^2 - (P_x-\frac{e}{c}A_X)^2 -
(P_Y-\frac{e}{c}A_Y)^2}\;\;.
$$
It can be proven that this is indeed a canonical transformation.
Note the existence of the reference orbit for zero vector potential, for which $(X,Y,P_X,P_Y, P_s) =(0,0,0,0,P_0) $.

\subsubsection{Neglecting electric fields}
 Due to the fact that longitudinal (synchrotron) motion is much slower than the transverse (betatron) one, the electric field can be set to zero and the Hamiltonian is written as
$$
{\cal \tilde H} = - \frac{e}{c}A_s- \left(1+\frac{X}{\rho(s)}\right)
\sqrt{(\frac{\cal {H}}{c})^2 - m^2 c^2 - (P_x-\frac{e}{c}A_X)^2 -
(P_Y-\frac{e}{c}A_Y)^2}\;\;.
$$
Using $\displaystyle P^2=\left( \frac{\cal {H}}{c}\right)^2 - m^2 c^2$, the Hamiltonian becomes
$$
{\cal \tilde H} = - \frac{e}{c}A_s- \left(1+\frac{X}{\rho(s)}\right)
\sqrt{(P^2 - (P_x-\frac{e}{c}A_X)^2 -
(P_Y-\frac{e}{c}A_Y)^2}\;\;.
$$
If static magnetic fields are considered, the time dependence is also dropped, and the system is having 2 degrees of freedom + “time” (path length $s$).

\subsubsection{Momentum rescaling}
 Due to the fact that the total momentum is much larger than the transverse ones, another transformation may be considered, where the transverse momenta are rescaled:
\begin{eqnarray*}({\bf Q},{\bf P}) & \mapsto & ({\bf \bar{q}},{\bf \bar{p}}) \;\; \text{or}\\
(X,Y,t,P_X,P_Y,P_t) & \mapsto & (\bar{x},\bar{y},\bar{t},\bar{p}_x,\bar{p}_y,\bar{p}_t) = (X,Y,-c \ t,\frac{P_X}{P_0},\frac{P_Y}{P_0},-\frac{P_t}{P_0c})\;\;.
\end{eqnarray*}
The new variables are indeed canonical if the Hamiltonian is also rescaled and written as 
$$
{\cal \bar{H}} (\bar{x}, \bar{y}, \bar{t}, \bar{p}_x, \bar{p}_y, \bar{p}_t)= \frac{\cal \tilde{H}}{P_0} = - e\bar{A}_s- \left(1+\frac{\bar{x}}{\rho(s)}\right)
\sqrt{\bar{p}_t^2 - \frac{m^2 c^2}{P_0} - (\bar{p}_x- e\bar{A}_x)^2 - (\bar{p}_y- e\bar{A}_y)^2}\;\;,
$$
 with $\displaystyle  (\bar{A}_x,\bar{A}_y,\bar{A}_s) = \frac{1}{P_0 \, c}(A_x,A_y,A_s)$  and $\displaystyle  \frac{m^2c^2}{P_0} = \frac{1}{\beta_0^2\gamma_0^2}$.

\subsubsection{Moving the reference frame}
Along the reference trajectory $\displaystyle {\bar p}_{t0} = \frac{1}{\beta_0}$ and $ \displaystyle \frac{d{\bar t}}{ds} \big|_{P=P_0} = \frac{\partial {\bar H}}{\partial {\bar p}_t} |_{P=P_0} = -\bar{p}_{t0} = -\frac{1}{\beta_0}$.
It is thus useful to move the reference frame to the reference trajectory for which another canonical transformation is performed
\begin{eqnarray*}({\bf \bar{q}},{\bf \bar{p}}) & \mapsto & ({\bf \hat{q}},{\bf \hat{p}}) \;\; \text{or}\\
(\bar{x},\bar{y},\bar{t},\bar{p}_x,\bar{p}_y,\bar{p}_t) & \mapsto & (\hat{x},\hat{y},\hat{t},\hat{p}_x,\hat{p}_y,\hat{p}_t) = (\bar{x},\bar{y},\bar{t}+\frac{s-s_0}{\beta_0},\bar{p}_x,\bar{p}_y,\bar{p}_t-\frac{1}{\beta_0})\;\;.
\end{eqnarray*}
The mixed variable generating function is $\displaystyle ({\bf \hat{q}},{\bf \bar{p}}) = (\frac{\partial F_2({\bf \bar{q}},{\bf \hat{p}})}{\partial {\bf \hat{p}}}, \frac{\partial  F_2({\bf \bar{q}},{\bf \hat{p}})}{\partial {\bf \bar{q}}} )$ providing
$$
F_2({\bf \bar{q}},{\bf \hat{p}}) = \bar{x}\hat{p}_x + \bar{y}\hat{p}_y + (\bar{t}+\frac{s-s_0}{\beta_0})(\hat{p}_t +\frac{1}{\beta_0})\;\;.
$$
The Hamiltonian is then
$$
{\cal \hat{H}} (\hat{x}, \hat{y}, \hat{t}, \hat{p}_x, \hat{p}_y, \hat{p}_t)=  \frac{1}{\beta_0}(\frac{1}{\beta_0}+\hat{p}_t) - e\hat{A}_s- \left(1+\frac{\hat{x}}{\rho(s)}\right)
\sqrt{(\hat{p}_t+\frac{1}{\beta_0})^2 - \frac{1}{\beta_0^2\gamma_0^2} - (\hat{p}_x- e\hat{A}_x)^2 - (\hat{p}_y- e\bar{A}_y)^2}\;\;.
$$

\subsubsection{Relativistic and transverse field approximations}
 First note that $\displaystyle  \hat{p}_t = \bar{p}_t-\frac{1}{\beta_0} = \bar{p}_t- \bar{p}_{t0} = \frac{P_t-P_0}{P_0} \equiv \delta$ and $l = \hat{t}$. In the ultra-relativistic limit $\displaystyle \beta_0 \rightarrow 1\;, \;\; \frac{1}{\beta_0^2\gamma^2}\rightarrow 0$ and the Hamiltonian is written as
 $$   
 {\cal {H}} ({x}, {y}, l, {p}_x, {p}_y, \delta)=  (1+\delta)- e\hat{A}_s- \left(1+\frac{{x}}{\rho(s)}\right)
\sqrt{(1+\delta)^2 - ({p}_x- e\hat{A}_x)^2 - ({p}_y- e\hat{A}_y)^2}\;\;,
$$														              
where the “hats” are dropped  for simplicity. If we consider only transverse field components, the vector potential has only a longitudinal component and the Hamiltonian is written as
$$
{\cal {H}} ({x}, {y}, l, {p}_x, {p}_y, \delta)=  (1+\delta)- e\hat{A}_s- \left(1+\frac{{x}}{\rho(s)}\right)
\sqrt{(1+\delta)^2 - {p}_x^2 - {p}_y^2}\;\;.
$$
Note that the Hamiltonian is non-linear even in the absence of any field component (i.e. for a drift)!

%
%
%
%
%
%
%
%
%


%
%
%
%
%
%

\end{document}